\documentstyle[12pt,aaspp]{article}

\lefthead{PAGE}
\righthead{GEMINGA: A COOLING SUPERFLUID NEUTRON STAR}

\newcommand{\Msol}{\mbox{$M_{\odot}\;$}}

\newcommand{\tP}{\mbox{$^{3}P_{2}\;$}}
\newcommand{\sS}{\mbox{$^{1}S_{0}\;$}}

\newcommand{\etal}{{\em et al. }}

\begin{document}

\title{Geminga: A Cooling Superfluid Neutron Star}
\author{Dany Page}
\affil{Department of Astronomy, Columbia University \\
       538 West 120th Street, New York, NY 10027}

\begin{abstract}

	We compare the recent temperature estimate for Geminga with
neutron star cooling models.
	Because of its age ($\sim 3.4 \times 10^5$ yr), Geminga is
in the photon cooling era.
	We show that its surface temperature
($\sim 5.2 \times 10^5$ K) can be understood by both types of neutrino
cooling scenarios, i.e, slow neutrino cooling by the modified Urca
process or fast neutrino cooling by the direct Urca process or by some
exotic matter, and thus does not allow us to discriminate between these
two competing schemes.
	However, for both types of scenarios, agreement with the observed
temperature can only be obtained if {\em baryon pairing is present
in most, if not all, of the core of the star}.
	Within the slow neutrino cooling scenario, early neutrino
cooling is not sufficient to explain the observed low temperature, and
extensive pairing in the core is necessary to reduce the specific heat
and increase the cooling rate in the present photon cooling era.
	Within all the fast neutrino cooling scenarios, pairing is
necessary throughout the whole core to control the enormous early
neutrino emission which, without pairing suppression, would result in a
surface temperature at the present time much lower than observed.

	We also comment on the recent temperature estimates for PSR
0656+14 and PSR 1055-52, which pertain to the same photon cooling era.
	If one assumes that all neutron stars undergo fast neutrino
cooling, then these two objects also provide evidence for extensive
baryon pairing in their core, but observational uncertainties also
permit a more conservative interpretation, with slow neutrino emission
and no pairing at all.
	We argue though that observational evidence for the slow
neutrino cooling model (the ``standard'' model) is in fact very dim and that
the interpretation of the surface temperature of all neutron stars could
be done with a reasonable theoretical {\em a priori} within the fast
neutrino cooling scenarios only.
	In this case, Geminga, PSR 0656+14, and PSR 1055-52 all show evidence
of baryon pairing down to their very centers.

\bigskip {\em The Astrophysical Journal}, in press.

\end{abstract}


\section{INTRODUCTION}

	The study of the thermal evolution of young neutron stars offers
the possibility of obtaining unique information about the structure of
compressed cold nuclear matter.
	Early cooling after the supernova explosion is driven by
neutrino emission, whose rate is a very sensitive function of the state
of that matter.
	Moreover, both the neutrino cooling and the later photon cooling are
strongly affected by the occurence of pairing ({\em \`{a} la} BCS) of
the baryonic components of the star's core.
	These two aspects combined result in a wide range of predicted
surface temperatures and give us two handles by which
to extract information about
the composition and pairing state of dense matter through comparison of
neutron star cooling calculations with observations of neutron stars of
known ages.

	Observational candidates for this purpose must have an age well
below $10^6$ yr since, after this, the star has exhausted its initial
heat content and its (much lower) temperature depends on other
mechanisms.
	Interstellar absorption is significant at the photon energies
corresponding to the expected surface temperatures, of the order of
$10^6$ K or lower, and hence the star must be not too far away from us
and must be in a region of low interstellar absorption.
	Until recently, only three neutron stars fullfilled these two
criteria of relative youth and closeness, all three located within the
local bubble of low interstellar matter density surrounding the Sun: PSR
0833-45 (Vela), PSR 0656+14, and PSR 1055-52.
	They provided the only
reliable data to compare with theoretical models.
	Some other objects, for example PSR 0531+21
(Crab; Harnden \& Seward 1984)
or the neutron star in the supernova remnant 3C58
(Becker, Helfand \& Szymkowiak 1982),
located at farther distances have
only given rough upper limits of their surface temperatures.
	We refer to \"{O}gelman (1991) for a review of the pre-{\em ROSAT}
observational situation.

	With the recent demonstration by Halpern \& Holt (1992)
that Geminga {\em is} a neutron star, a new candidate is now available.
	The quality of the {\em ROSAT} data makes
this one of the best cases of
detection of thermal radiation from a neutron star surface to date,
and the data analysis (Halpern \& Ruderman 1993)
is the most detailed performed so far.
	Geminga is one of the closest neutron stars and is located
within the local interstellar bubble;
	Gehrels \& Chen (1993) recently argued that the Geminga
supernova may have actually been the cause of this bubble.

	Comparison of Geminga's age and temperature with published
models of neutron star cooling shows that the data can be accomodated by
a variety of models.
	Several of the direct Urca cooling scenarios of Page \&
Applegate (1992), the kaon condensate cooling scenario of
Page \& Baron (1990), and some of the pion condensate cooling
scenarios of Umeda {\em et al.} (1992) work,
while none of the fast cooling models, either with
quarks or pion condensate, of Van Riper (1991) is successful.
	Some, but not all, of the ``standard'' cooling models presented by
Nomoto \& Tsuruta (1986, 1987), Shibazaki \& Lamb (1989),
Page \& Baron (1990), Van Riper (1991),
Page \& Applegate (1992), and Umeda {\em et al.} (1992, 1993) are also
successful.
	In this paper, we will look at the various ingredients of these
models and determine which ones are crucial for compatibility with this
new neutron star.
	A preliminary version of this work has been presented in
Page (1992).

	The pulsars PSR 0656+14 and PSR 1055-52 have ages similar to
Geminga's and fit into the study of the present work.
	Since the analyses of the {\em ROSAT} observations of these
two objects have been published recently (Finley, \"{O}gelman \&
Kizilo\mbox{\u{g}}lu 1992; \"{O}gelman \& Finley 1993) we will also
discuss them briefly.

	The structure of the paper is as follow.
	Section 2 presents the observational data on Geminga.
	Section 3 describes the general physics of neutron star cooling
relevent to our present purpose.
	Section 4 discusses the various fast neutrino cooling scenarios,
and section 5 presents detailed calculations within the slow neutrino
cooling scenario.
	Comparison with Geminga is done in sections 4 and 5 while section
6 comments on the relevence of the previous results for other pulsars,
in particular for PSR 0656+14 and PSR 1055-52.
	Section 7 contains our conclusions.

\section{GEMINGA}

\subsection{Geminga's Age}

	The Geminga period as measured by Halpern \& Holt (1992)
is $P = 0.2370974 \; s \pm 0.1 {\mu}s$.
	The earlier observation by {\em COS-B} and the later ones by {\em GRO}
 have slightly
different $P$s which lie very accurately on a straight line
(Bignami \& Caraveo 1992)
and give a practically constant period derivative of
$\dot{P}=1.099 \pm 0.001 \times 10^{-14} \; s \; s^{-1}$
over a span of 16 yr.
	The corresponding spin-down age is $\tau = P/2\dot{P} = 3.4
\times 10^5$ yr.
	This age is obtained with a braking index $n = 3$, i.e.,  with
magnetic dipole braking.
	We will consider a range of ages corresponding to braking
indices $n$ = 2 and 4,
\begin{equation}
2.3 \times 10^5 {\rm yr} \leq t \leq 6.8 \times 10^5 {\rm yr},
\end{equation}
the upper value probably being an overestimate.
	We refer to Michel (1991) and Lyne \&
Graham-Smith (1990) for discussions of the reliability of the spin-down
age as an indicator of the true age.

\subsection{Geminga's Temperature}

	In the first estimate of the star's temperature, Halpern
\& Holt (1992) fitted the two apparent components of the spectrum with a
blackbody and a power law.
	The best fit gave a blackbody temperature of $T = 3 -4 \times
10^5$ K.
	In a second, more detailed analysis Halpern \& Ruderman (1993)
replaced the power-law component by a second blackbody with higher
temperature and argued that this hotter component ($T \cong 3 \times
10^6$ K) is due to emission from a reheated polar cap (this temperature
is too high to be explained only by the anisotropy of heat transport in
the crust in presence of a magnetic field).
	They obtained for the main surface emission a temperature $T =
5.2 \pm 1.0 \times 10^5$~K.
	From the luminosities of these two components, they
concluded that the ratio of the areas of the hot and cold emitting
regions is about $3 \times 10^{-5}$.
	Moreover, it is likely that the surface is not emitting uniformly
and that some colder region is not being seen, making the concept of
``surface temperature'' an ambiguous one.
	However, cooling calculations give as an output the effective
temperature, i.e.,  a total luminosity, and the presence of a cooler
region would reduce the luminosity and make the effective temperature
somewhat lower than the inferred value of $5.2 \pm 1.0 \times 10^5$ K.

	All these analyses use blackbody spectra, but the surface of a
neutron star cannot be expected to be a perfect blackbody.
	Romani (1987) has calculated more realistic spectra
for various surface chemical compositions without magnetic field.
	Miller (1992) and Shibanov {\em et al.} (1992)
have partially extended these results by including the magnetic field effects.
	The general trend of these results, for H or He atmospheres,
is that there is an excess emission in the Wien tail of the spectrum,
compared to a blackbody, and the excess falls within the {\em Einstein} and
{\em ROSAT} detector ranges.
	This excess is reduced if metals are present (because of absorption
edges) or when the effects of the magnetic field are taken into account.
	As a consequence, the use of these spectra would lower the
measured temperature.
	Finally, some contamination from a surrounding nebula and/or
some surface reheating by gamma rays or particles from the magnetosphere
cannot be excluded.
	See Halpern \& Ruderman (1993) for a discussion.

	We will take for comparison with our calculations an effective
temperature of
\begin{equation}
4 \times 10^5 {\rm K} \leq T_e \leq 6 \times 10^5 {\rm K}
\end{equation}
but insist that it must be taken as an upper limit for the above
mentioned reasons.
	Being determined from the spectrum, this temperature is
independent of the distance, mass, and radius of the star and is the
``temperature at infinity'' $T^{\infty}$, i.e.,  the redshifted
temperature.

\pagebreak
\section{THE PHYSICS OF NEUTRON STAR COOLING}

	At an age of a few multiples of $10^5$ yr {\em Geminga is
well into the isothermal phase} (Nomoto \& Tsuruta 1987);
	its internal temperature $T_i$ is uniform except for a gradient
in the {\em envelope} just below the surface.
	The thermal evolution of the star is determined by energy
conservation,
\begin{equation}
\frac{dE}{dt} = C_v \frac{dT_i}{dt} = - L_{\nu} - L_{\gamma},
\end{equation}
where $E$ is the total thermal energy of the star,
$C_v$ its total specific heat, and $L_{\nu}$ and $L_{\gamma}$ the total
neutrino and photon luminosities (general relativistic correction
factors are omitted here but were included in our calculations).
	The photon luminosity is $L_{\gamma} = 4 \pi R^2 \sigma T_e^4 \;
{\propto} \; T_i^{2.2}$, where the effective temperature $T_e$ is
converted into an internal temperature $T_i$ according to the $T_i - T_e$
relationship calculated by Gudmundsson, Pethick \& Epstein
(1982, 1983).
	The neutrino emissivities are proportional to $T_i^6$ or $T_i^8$
(see Table \ref{tab:nu}), hence photon emission from the surface will
eventually dominate over neutrino emission when the temperature has
dropped sufficiently.
	This turnover happens before the star reaches $10^5$ yr :
{\em Geminga is thus well into the photon cooling era}.
	In this phase the cooling rate is mostly determined by the total
specific heat of the star, but the actual surface temperature also depends on
the previous neutrino cooling, which can be considered as giving the
initial condition for photon cooling.
	Our cooling curves are calculated with the Heyney-type code
already presented in Page \& Baron (1990) and Page \& Applegate (1992)
which solves the general relativistic heat transport and energy balance
equations.
	The reader is refered to these papers for more details.

\subsection{Neutrino Processes}

	The dominant neutrino emission processes occur in the
core of the star and are variants of beta and inverse beta decay.
	Table \ref{tab:nu} shows the approximate emissivities of several
processes for comparison.
	One can divide them into slow and fast neutrino emission,
according to wether they involve four or two baryons.
	The large difference between slow and fast processes comes mainly
from phasespace considerations: Fermi's ``Golden rule'' tells us that the
rate is proportional to the total phasespace volume available for both
initial and final particles, and for fermions this volume is proportional
to $k_B T/E_F$ where $E_F$ is the particle Fermi energy.
	Typical nucleon Fermi energies in neutron star cores are of the
order of a few tens to a few hundreds of MeV; if one takes a typical
temperature of $T = 10^9 {\rm K} \cong 0.1 {\rm MeV}/k_B$ and $E_F \sim
10^2$ MeV then a four fermion process is about $(k_B T/E_F)^2 \sim
10^{-6}$ time weaker than a two fermion process, which is what the
direct Urca - modified Urca emissivities show (see Table 1).
	Participation of a meson (pion or kaon) in a process does not
introduce any phase space limitation since these are bosons, but strong
interactions effects, and strangeness violation in the case of kaons,
reduce the efficiency of meson processes
compared to the simple direct Urca process.
	Hyperons may also constitute a large fraction of the core
baryons (see, e.g.,  Glendenning 1985) and will then participate into
either modified (Maxwell 1987) or direct (Prakash et al. 1992) Urca
processes, depending on their relative concentrations, but with somewhat
lower emissivities compared to the corresponding nucleon processes.
	We refer to Pethick (1992) and Prakash (1993)
for recent reviews of the neutrino emission problem.
	While there is still doubt about which fast process can actually
occur, the number of presently proposed channels is so large that it is
becoming difficult to believe that none of them is permitted and that
neutron star cooling follows the old ``standard'' model with only the slow
modified Urca process.
	Nevertheless, awaiting a conclusive argument on this point, we
will still consider both the fast and slow cooling scenarios.

	Deconfined quarks may be present in the center of massive
neutron stars and are also copious neutrino emitters.
	They thus belong to the fast neutrino cooling scenario but obviously
require a separate treatment.
	We will not consider them explicitly here.

\subsection{The Boundary Condition}

	An important ingredient is the above mentioned $T_i - T_e$
relationship.  Gudmundsson \etal (1982, 1983) were the
first to present a detailed study of it but did not include the effect
of the magnetic field $\vec{H}$ which enhances the heat transport
parallel to $\vec{H}$ and strongly suppresses it in the perpendicular
direction.
	The extensive magnetic envelope calculations of Hernquist (1985)
and Van Riper (1988) for transport parallel to $\vec{H}$ show that for a
given inner temperature $T_i$ the surface temperature $T_s$ is raised,
compared to the nonmagnetic case, but by no more than 50\% even with a
field of $10^{14}$ G (the surface magnetic field of Geminga is estimated
to be about $1.6 \times 10^{12}$ G).
	When the field is at an angle to the surface, $T_s$ should be
lower, and by simple geometric considerations Hernquist (1985) argued
that the magnetic effects when including a global field configuration
are probably very small.
	Schaaf (1990a,b) has performed envelope calculations with an
arbitrary orientation of $\vec{H}$ with respect to the surface, and his
results can be used to estimate the surface temperature distributions
$T_s(\theta,\phi)$ resulting from various magnetic field configurations
and the corresponding effective temperatures $T_e$.
	Preliminary results (Page 1994) confirm Hernquist's point that
the $T_i - T_e$ relationship depends only weakly on the magnetic
field.
	We will consequently use the zero-field relationship here, which
should introduce an error of no more than a few percent.

\pagebreak
\subsection{The High-Density Equation of State}

	The equation of state (EOS) has two functions in our models, the
first being simply to give the global structure of the star, i.e.,  the
density versus radius profile as a solution of the Oppenheimer-Volkoff
equation of hydrostatic equilibrium, and the second being to determine
the chemical composition.
	Obviously, the slow and fast neutrino cooling cases are associated with
quite different EOSs at supranuclear density, and we will discuss them
separately.

	For modeling the slow neutrino cooling, we consider five different
EOSs from modern calculations: one relativistic, MPA (M\"{u}ther, Prakash \&
Ainsworth 1987), two nonrelativistic, FP (Friedman \& Pandharipande 1981) and
WFF(av14) (Wiringa, Fiks \& Fabrocini 1988), and two parametric,
PAL32 and PAL33 (together PAL: Prakash, Ainsworth \& Lattimer 1988), whose
properties are intermediate to the above three.
	Each one of these calculations gives the energy per baryon for
neutron matter and symmetric nuclear matter, from which we obtain the EOS
for matter in $\beta$-equilibrium and the chemical composition following
Wiringa \etal  (1988), or Prakash \etal (1988) for PAL.
	The properties of these EOSs related to neutron stars are
summarized in Table \ref{tab:EOS};
	they encompass a broad range of stiffness with maximum masses
of 1.68 \Msol - 2.44 \Msol, proton fractions in a 1.4 \Msol star
of 2.0\% - 11.4\% and radii of a 1.4 \Msol star of 10.6 km -  12.5 km.
	We have rejected EOSs with proton fractions large enough to
allow the direct Urca process in a 1.4 \Msol star, but do consider
MPA, PAL32, and PAL33 which allow it at higher masses (notice that
the PAL32 EOS gives a 1.4 \Msol star on the verge of allowing the direct
Urca process).
	Two EOSs, BPS (as listed in Baym, Pethick \& Sutherland 1971)
and PS (Pandharipande, Pines \& Smith 1976) have been very popular in
neutron star cooling studies.
	We do not use the BPS EOS for modeling the slow cooling since
this extremely soft EOS (which gives a density above 10 times nuclear
matter density in the center of a 1.4 \Msol star) is actually built on a
high-density EOS with hyperons (Pandarhipande 1971) in which the
hyperonic direct Urca is allowed and this EOS thus belongs to the fast
neutrino cooling case (hyperons appear at a star mass of 0.4 \Msol in
this model).
	The softening of this EOS is due in an essential way to the
presence of the hyperons, and nothing similar can be expected with
only nucleons.
	The PS EOS has already been considered in previous works (Nomoto
\& Tsuruta 1986, 1987; Van Riper 1991), and we will simply quote their
results below.
	However, in this extremely stiff EOS the neutrons form a
three-dimensional lattice and thus have a totally different specific heat and
a different neutrino emissivity than liquid neutrons, two facts not taken
into account in the models, and the stiffness of this EOS is inseparable
from to the lattice structure.

	The choice of the EOS for fast neutrino cooling models is not as
important as for slow cooling at the present stage of development of
the theory.
	There are several other factors which have much more influence,
for example the critical density at which fast neutrino emission
turns on, the fast neutrino emission which is actually at work, the occurence
of baryon pairing, etc., all of which are at best poorly known.
	Therefore, calculations of fast cooling have often been done with
various EOSs in a partially justified, careless way.
	However, one model has been developed with much detail, based on the
ALS model of dense matter (Takatsuka {\em et al.} 1978), which is somewhat
inspired by the PS EOS, but where the lattice structure is one-dimensional
and where two-dimensional nucleon pairing occurs in planes orthogonal to the
lattice direction (see Tamagaki 1992 for a general description).
	In this model, charged pion condensation develops at high density,
but the resulting EOS is close to the FP EOS at low density below the
condensation threshold, i.e.,  the stiffness of the PS EOS has
disappeared.
	Cooling calculations within the ALS model have been performed by
Umeda {\em et al.} (1992), and their results will be used below.

\subsection{Nucleon Pairing} 

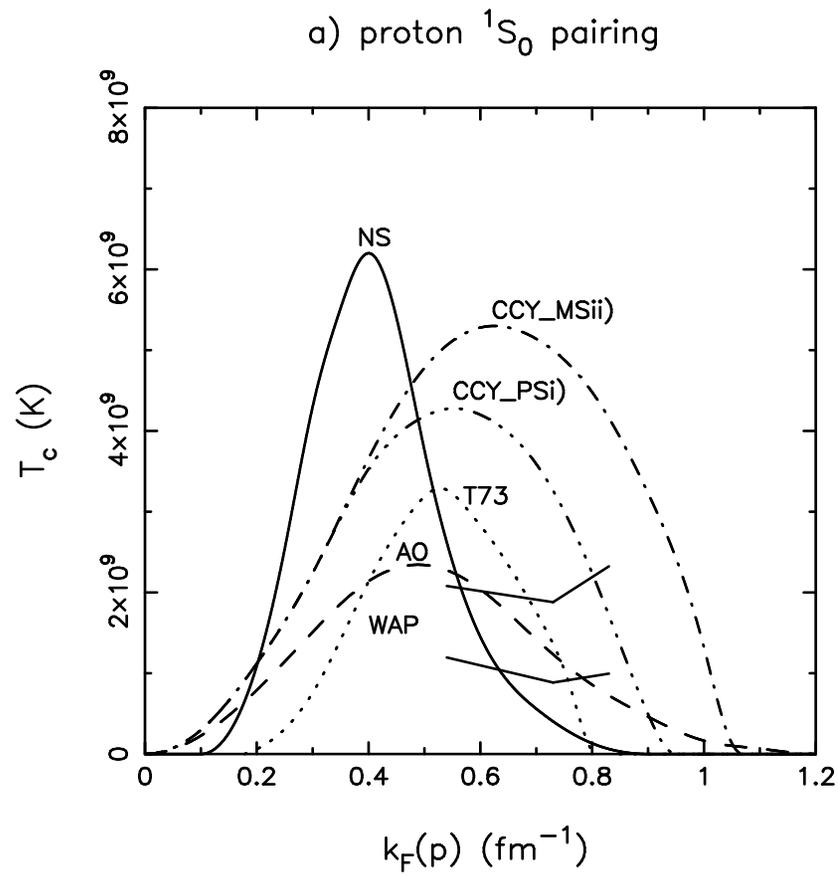
\begin{figure}
\caption{\label{fig1}
         a) Proton \sS pairing critical temperatures.
            CCY-Chao, Clark \& Yang (1972),
            T73-Takatsuka (1973),
            NS-Niskanen \& Sauls (1981),
            AO-Amundsen \& \mbox{\O}stgaard (1985a),
            WAP-Wambach \etal(1991).
     \newline
         b) Neutron \tP pairing critical temperatures.
            HGRR-Hoffgerg \etal (1970),
            T72-Takatsuka (1972),
            AO-Amundsen \& \mbox{\O}stgaard (1985b).
            The two dashed curves show the results of T72 and AO when the
            neutron effective mass is fixed at the free mass value.
     \newline
         In abscissa is the Fermi wave number $k_F$, related to the
         particle number density $n_i$ by
         $k_F(n_i) = (3 \pi^2 n)^{1/3} = 1.68 (n_i/n_0)^{1/3} {\rm fm^{-1}}$
         where $n_0 = 0.16 {\rm fm^{-3}}$ .}
\end{figure}

	Pairing {\em \`{a} la} BCS (superfluidity) of the nucleons in
the neutron star core has a dramatic effect on the cooling because it
suppresses both the neutrino emission and the specific heat.
	The pairing is in the \sS partial wave at low density and then
shifts to the \tP partial wave at higher density.
	The protons in the core and the neutrons in the inner crust
are expected to be paired in the \sS partial wave while the neutron
pairing shifts to the \tP partial wave in the core.
	The pairing of protons in the \tP partial wave seems never to have been
considered, probably on the grounds that its critical temperature would
be very low.
	At still higher densities, the pairing should shift to the
$^{1}D_{2}\;$ partial wave (the next partial wave in which the free
nucleon-nucleon interaction is attractive), but the estimated corresponding
critical temperature is too low for this type
of pairing to be of any interest (Amundsen \& \mbox{\O}stgaard 1985b).
	Theoretical calculations of $T_c$ are extremely difficult and
the presently published values are still highly uncertain except in the
case of crust neutron \sS pairing, where reasonable agreement has been
obtained between the various latest calculations.
	Figure~\ref{fig1} shows most of the presently published calculations
of critical temperatures for core neutron and proton pairings.
	One sees that both the maximum value of $T_c$ and the
density range where it is nonzero are very uncertain, particularly in
the neutron case.
	For the \sS proton pairing, the latest calculation (Wambach,
Ainsworth \& Pines 1991) was the first to include in a consistent
way the neutron background and shows a strong {\em reduction} of $T_c$,
but with a density dependence apparently different from that found
in earlier calculations.
	However, this result depends strongly on the relative densities
of neutrons and protons, and thus, calculations with a different proton
fraction (a poorly known quantity) may give quite different results
(Ainsworth 1992).
	Medium dispersion effects (i.e., change in the nucleon
effective mass) have an enormous effect, as shown in
Figure~\ref{fig1}b by the differences between the curves AO and T72,
where the effective mass $m^* \equiv M^*/M$ is obtained in a self-consistent
way and the corresponding curves labeled with $m^* =1$ for
which the effective mass is forced to the free mass value.
	(The Hoffberg {\em et al.} 1970 calculation used $m^* = 1$.)
	Of course, $m^*$ is also poorly known at high density.
	Further extensive calculations within the ALS model (Takatsuka
\& Tamagaki 1982, and references therein) with and without pion
condensation give values of $T_c$ for \tP neutron pairing that are between the
two extremes shown in Figure~\ref{fig1}b as T72 and T72($m^* = 1$).
	For \tP neutron pairing, background effects beyond the first
order ones considered in the results shown in Figure~\ref{fig1}b and similar to
the effects considered by Wambach \etal (1991) for \sS
proton pairing have been studied (Jackson {\em et al.} 1982).
	The results indicate that, in this case, background effects
strongly {\em enhance} the pairing; thus, values of $T_c$ higher than
shown in Figure~\ref{fig1}b and extending to higher densities are quite
possible.
	In light of this, the case for proton \tP pairing should also be
considered seriously, as should $^{1}D_{2}\;$ pairing.
	In other words, we know neither the value of $T_c$ in the core
nor the relevent density range, neither for neutrons nor for protons, but
it is reasonable to expect large values extending to high densities.
	If hyperons are present, one can expect that they will also
pair for the same reasons nucleons do.
	With regard to quarks,
pairing is also very probable (Bailin \& Love 1984).

	The effect of pairing is to reduce the phase space available for
excitations.
	As a result, both the specific heat of the paired component and
the neutrino processes in which it participates will be suppressed.
	We treat this effect on $C_v$ following Levenfish \& Yakovlev
(1993; see also Gnedin \& Yakovlev 1993), who performed detailed
calculations of the reduction factors in cases both of isotropic \sS and
anisotropic \tP pairing.
	For the neutrino emissivity suppression, we simply use a Boltzmann
factor $\exp(- \Delta / kT)$, where $\Delta$ is the pairing gap, which
is not very accurate but has no serious consequence here since we
consider only the photon cooling era.

\section{FAST NEUTRINO COOLING SCENARIOS}

	Fast neutrino cooling encompasses a variety of different
scenarios, kaon or pion condensate, direct Urcas with nucleons and/or
hyperons and/or isobars, quark matter, and so forth which share the
characteristic that their neutrino emissivities are many orders of
magnitude higher than the modified Urca process emissivity.
	As a result, whenever one of these processes is allowed to
operate freely, the resulting surface temperature, once the star has
reached isothermality, is far below any of the presently available
observational estimates (Page \& Baron 1990; Van Riper 1991; Page \&
Applegate 1992; Page 1992; Umeda {\em et al.} 1992), including the new
Geminga observation.
	This raw picture is strongly altered if the neutrino-emitting
baryons (nucleons, hyperons, quarks, etc.) become paired.
	As presented by Page (1989) for the kaon condensate case and
by Page \& Applegate (1992) for the direct Urca with nucleons case, the early
suppression of the neutrino emission due to the pairing gap can keep the
surface temperature $T_s$, after isothermalization and until an age of
about $10^5$ yr, anywhere between approximately $2 \times 10^5$ K and
$1.5 \times 10^6$ K.
	The actual value of $T_s$ is then only a function of the pairing
critical temperature $T_c$ of the neutrino emitting fluid(s) (to be
precise, since $T_c$ is density dependent, $T_s$ is a function of the
lowest value of $T_c$ within the ``pit'' of fast neutrino emission).
	In a multicomponent system like nucleons + hyperons, various
direct Urca processes can operate simultaneously (Prakash et al.  1992),
and, for the star not to drop into invisibility, all of them have to
be stopped by having one of the participating baryons paired .
	In particular, if both $\Lambda$ and $\Sigma^{-}$ are present
together, they undergo a purely hyperonic direct Urca process, and thus,
one of them must be paired.
	One can expect that $T_c$ is lower for hyperons than for
nucleons if the former have weaker interactions than the latter, and the
star's temperature would then be controlled by hyperon pairing.

	The models of Page \& Baron (1990), Umeda {\em et al.}
(1992), and Page \& Applegate (1992) of kaon, pion, and nucleon direct
Urca cooling, respectively, with superfluidity suppression easily
accomodate the estimated surface temperature of Geminga.
	The crucial point, however, is that pairing has to occur down to
the very center of the star and the critical temperature must be higher
than $10^9$ K everywhere.
	If even a very small region is left unpaired, it will drive the
surface temperature well below the observed value of $5 \times 10^5$ K.
	For example, the 1.4 \Msol case of Page \& Applegate (1992), with
a direct Urca emitting pit of only 0.038 \Msol, has a temperature of $1.5
\times 10^5$ K at Geminga's age if pairing does not occur.
	Such a low surface temperature would make Geminga's surface
practically unobservable by {\em ROSAT} (the hot polar cap would of course
still be seen).

	It is not possible to distinguish between the various neutrino
emission processes from this analysis only:
	both a kaon (or pion) condensate and the direct Urca, even if
their emissivities differ by three orders of magnitude, are compatible
with the Geminga data when pairing is taken into account, but with
different values for $T_c$.
	The $T_c$ values needed are within the range of theoretical
predictions, but this range is so large it can accomodate almost
any data.
	Moreover, even within one given neutrino emission scheme very
different values of $T_c$ are possible:
	for example the models labeled HGRR and 0.1HGRR of Fig. 2 in Page \&
Applegate (1992), i.e.,  direct URCA and \tP neutron pairing with $T_c$
from Hoffberg {\em et al} (1970) or with the same $T_c$ multiplied by 0.1,
are both acceptable within the observational uncertainty in the Geminga
data.

	It should be mentioned that internal heating by friction of the
crust neutron superfluid can significantly alter the thermal evolution
of a neutron star when its temperature, and thus its specific heat, is low.
	The models of fast cooling with pions and heating of Umeda {\em
et al.} (1993), without core nucleon pairing, can produce a star with a
surface temperature at Geminga's age of at most $3 \times 10^5$ K {\em
when heating is at its maximum strength} (instead of $1.5 \times 10^5$ K
without heating).
	This is lower that the Geminga temperature considered here
(measured with a blackbody spectrum) but may be high enough if Geminga
is actually cooler.

 \section{THE SLOW NEUTRINO COOLING SCENARIO}

	Slow neutrino cooling is a well-defined scenario
based on the conservative hypothesis that the neutron star core is made
exclusively of neutrons and protons (plus electrons and muons to
preserve charge neutrality, but no charged pions, kaons, quarks, etc.)
with a proton fraction low enough for the direct Urca process to be
forbidden (Lattimer {\em et al.} 1991), and its predictions for surface
temperatures are much more restrictive than those of the fast neutrino
cooling scenarios.
	We defer a detailed study to later work and only analyze here
the photon cooling era relevant to Geminga.

	The photon energy loss can be calculated accurately since
the core temperature-effective temperature relationship is known quite
accurately, even in presence of a magnetic field.
	We take the core neutrino emissivity of the modified Urca
process and the two associated neutral current bremstrahlung processes
from Friman \& Maxwell (1979).
	Of critical importance here is the total specific heat of the
star, which depends on the EOS and chemical composition (proton
fraction).
	Pairing of nucleons is essential here because of its
suppression of the specific heat.
	Table \ref{tab:Cv} shows the contribution to the normal (i.e.,
without pairing) specific heat of the various components in a $1.4 \;
M_{\odot}$ star at a temperature $T = 10^9 K$ for our five EOSs.
	Pairing will suppress $C_v$ exponentially when $T \ll T_c$ and
the corresponding specific heat will practically disappear.
	In all our calculations, the crust neutrons are paired using the
gap from Ainsworth, Wambach \& Pines (1989) and their contribution to
$C_v$ is thus strongly reduced; the crust electrons make a negligible
contribution as does the crust lattice.
	One can see from Table 3 that the core neutrons contribute
about three-fourths of the total specific heat, the protons
one-fourth and the core electrons about 5\%, independent of the
EOS.

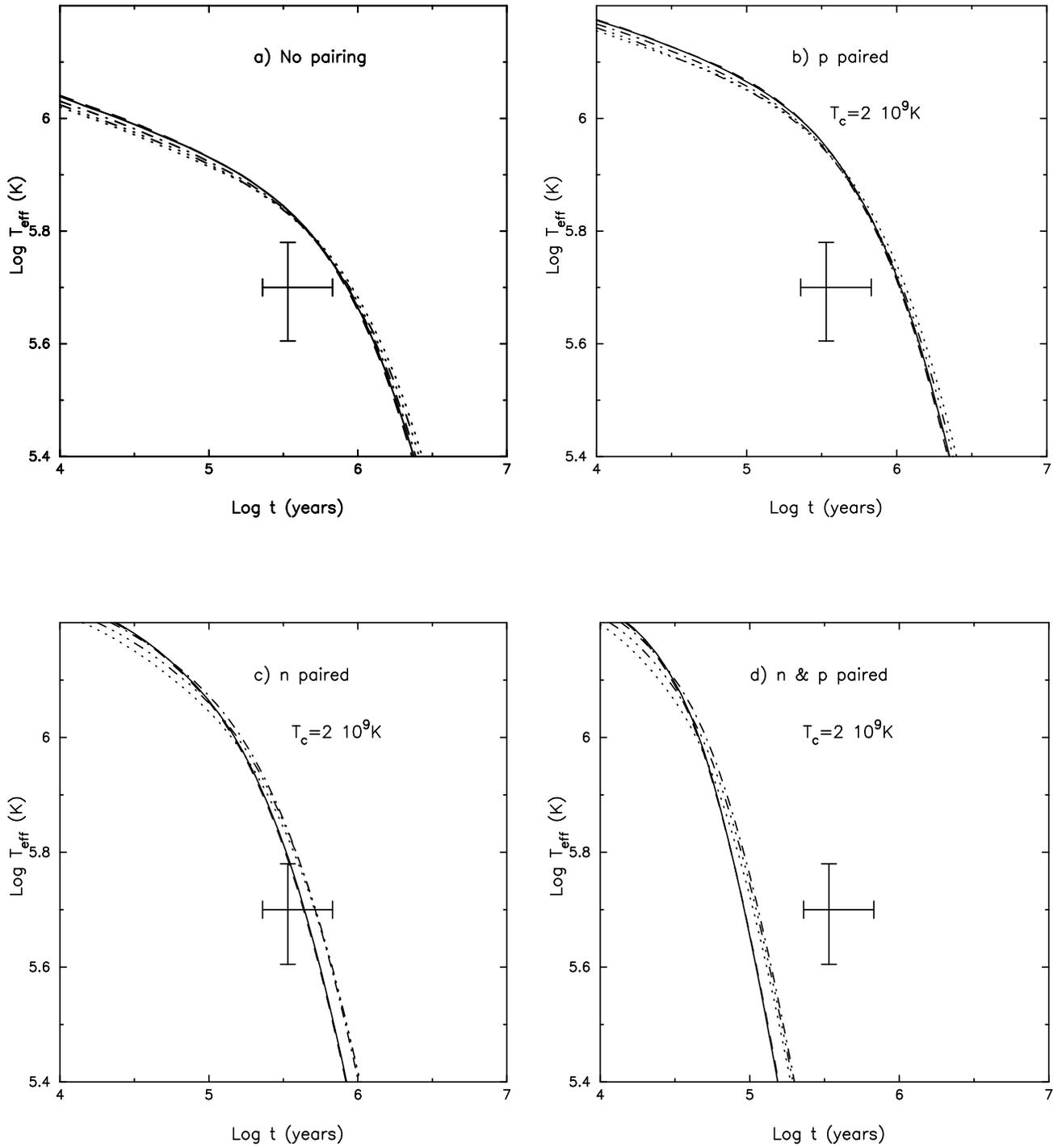
\begin{figure}
\caption{\label{fig2}
         Cooling by the modified Urca process:
         effect of the specific heat suppression by nucleon pairing.
         The various curves correspond to the five EOSs we use:
         FP (continuous), WFF(av14) (dashed), MPA (dotted),
         PAL32 (dash-dotted) and PAL33 (dash-triple-dotted).
         1.4 \Msol star.
\newline
         a) No core pairing at all.
\newline
         b) Protons paired with a density-independent
            $T_c = 2 \times 10^9$ K.
\newline
         c) Core neutrons paired with a density-independent
            $T_c = 2 \times 10^9$ K.
\newline
         d) Protons and core neutrons paired with a density-independent
            $T_c = 2 \times 10^9$ K.
         Crust neutrons are paired according to Ainsworth \etal (1989).
         The temperature in ordinate is the effective temperature
         ``at infinity,'' i.e., redshifted.
         The cross shows Geminga's temperature and age.}
\end{figure}

	Because of the large theoretical uncertainty about the actual value of
the pairing critical temperature $T_c$ for both neutrons and protons in
the core as well as the uncertainty on the density dependence of $T_c$
we first consider density independent $T_c$; i.e., we force pairing in
the whole core and take $T_c = 2 \times 10^9$ K.
	The resulting cooling curves of a 1.4 \Msol star for our five EOSs
are shown in Figures~\ref{fig2}a-\ref{fig2}d and are compared with the Geminga
data.
	Because in this photon cooling era the determining factor is the
total specific heat, the five EOSs give basically identical results
since the contribution of the various components to $C_v$ is only weakly
dependent on the EOS and pairing, when assumed, is present in the whole
core.
	When none of the core nucleons is paired (a) the theoretical
results are consistent only with the higher surface temperature $T_s$
and the older age:
	considering that this $T_s$ is certainly an
overestimate, one can state that Geminga's temperature is incompatible
with these cooling models (unless Geminga's age is underestimated,
but spin-down ages are usually considered to be overestimates).
	With pairing of the protons (b) the discrepancy increases
because the small ($\sim 25\%$) decrease of the specific heat in the
photon cooling era is not sufficient to compensate for the significant
reduction of the earlier core neutrino emission (during which only the
very slow $nn$ bremsstrahlung is unaffected), which gives a high
temperature at the beginning of the photon cooling era.
	When all the core neutrons are paired (c) the reduction of $C_v$
is large enough to accomodate the observed temperature.
	If both neutrons and protons are paired within the whole core
(d) $C_v$ is cut by a factor of 20 (only the electron and muon contributions
are left)
and the temperature drop during the photon cooling era is extremely
fast; however, heating mechanisms (Shibazaki \& Lamb 1989; Cheng et al.
1992; Umeda {\em et al.} 1993) could slow the cooling and keep the
temperature compatible with Geminga.
	These results were at a fixed mass of 1.4 \Msol, but Figure~\ref{fig3}
shows that changing the star mass makes little difference as long as
pairing is still assumed throughout the whole core.
	With realistic density dependent gaps, by varying the star's mass
we change the fraction of core baryons paired, and any temperature
between the extreme cases of Figures~\ref{fig2}b and \ref{fig2}d could
in principle be obtained.

	Figure~\ref{fig4} shows three cooling curves with three published
calculations of \tP neutron pairing and the EOS WFF(av14) for a 1.4
\Msol star.
	(The Fermi momentum of the neutrons in the center of the star,
for comparison with Fig.~\ref{fig1}, is $k_F(n) = 2.58 fm^{-1}$).
	The pairing of Takatsuka (1972) has almost no effect, since $T_c$
vanishes at low density and most of the core is left unpaired.
	The two calculations of Hoffberg {\em et al.} (1970) and
Amundsen \& \mbox{\O}stgaard (1985b) give almost identical results in the
photon cooling era, since in both cases the whole core is paired and the
internal temperature $T_i$ is much lower than $T_c$;
	in both cases, the whole core neutron specific heat has been
practically eliminated.
	In the neutrino cooling era where $T_i$ is higher these two
cases do differ significantly (their $T_c$ differ by an order of magnitude):
	in the Hoffberg {\em et al.} (1970) case, the core neutrino
emission is practically turned off early on due to the very high $T_c$,
while in the Amundsen \& \mbox{\O}stgaard (1985b) case, the suppression
happens much later and is thus less efficient.
	It is thus not possible to deduce any value for $T_c$ from the
Geminga data only, except that it must be higher than a few multiples of
$10^8$ K, but we can state that pairing must occur within most of the
core.
	Published cooling curves with the PS EOS (Nomoto \& Tsuruta
1986, 1987; Van Riper 1991) also fit the Geminga data, in their superfluid
versions where the neutrons are paired in the whole core.

\section{COMMENTS}

\subsection{The Cooling of Low-Mass Neutron Stars}

	The surface temperatures we have obtained with the slow
neutrino cooling scenario during the photon cooling era when both
neutrons and protons are paired within the whole core (Fig.~\ref{fig2}d) are
much lower than any prediction previously published.
	This is so because 95\% of the star's specific
heat has been suppressed by superfluidity when all core nucleons are
paired.
	This case has to be seriously considered for low-mass neutron
stars, where there is little doubt that pairing happens down to the
center of the star for both neutrons and protons, and it has some
unexpected effects.
	Consider, e.g. as shown in Figure~\ref{fig5}, the case of a
heavy star undergoing fast neutrino cooling with suppression from
neutron \tP pairing but with a core that still has a substantial amount of
unpaired protons.
	If the critical temperature for neutron pairing is a few multiples of
$10^9$ K, during the neutrino cooling era the star will have a lower
temperature  than a slow neutrino cooling star of low mass, but later
during the photon cooling era it will be much warmer than the lighter
(wholly paired) star.
	Thus {\em fast neutrino cooling does not mean fast cooling
forever}.

	If one adopts the idea that the critical density for the onset
of fast neutrino emission is low, then it is quite possible that all
neutron stars undergoing slow neutrino cooling, and thus having very low
central densities, have both their neutrons and their protons paired within
the whole core with high values of $T_c$.
	All neutron stars undergoing slow neutrino cooling would then
cool very quickly in the photon cooling era and would become invisible
after a couple of hundreds of thousands of years.
	Consequently, any neutron star older than that, with detectable
surface thermal emission, has an unpaired baryonic core component, which
provides the star with a sizable specific heat, but has undergone fast
neutrino emission suppressed early on by pairing of its other core
component(s).

\pagebreak
\subsection{``Standard'' Cooling ?}

	However, there may not even be such a thing as a slow neutrino
cooling (``standard'' cooling) neutron star since, as mentioned in the
introduction, the number of presently proposed channels for fast
neutrino emission is so large, and the corresponding critical densities
so low in some cases, that it is possible that {\em all} neutron stars
cool by some fast neutrino process.
	If this is the case, the thermal evolution of all neutron stars
is entirely controlled by superfluidity.
	The surface temperature from age $\sim 10^2$ yr to $\sim
10^5$ yr depends on the minimal value of $T_c$ for the relevent
neutrino emitting baryonic component (nucleon or hyperon) in the core
(Page \& Applegate 1992), and the temperature from age $\sim
10^5$ yr to $\sim 10^6$ yr depends on how much of the core is
left unpaired.
	It is worth mentioning here that observational support for
the ``standard'' model is actually extremely dim, if not nonexistent.
	It has traditionally been based (Tsuruta 1986) on {\em Einstein}
observations of the Crab pulsar (Harnden \& Seward 1984) and of the two
compact objects detected in the supernova remnants 3C58 (Becker \etal 1982)
and RCW 103 (Tuohy {\em et al} 1983), all three
having ages $\sim$ 1000 yr and upper limits on temperature of the
order of $2-3 \times 10^6$ K.
	However, none of these three temperature estimates can be given
much credibility for the following reasons:
	1) {\em ROSAT} has failed to detect the previously seen source
in RCW 103 (Becker {\em et al} 1993); no matter what this object is, or
was, it is not a neutron star cooling according to the slow neutrino
emission scenario.
	2) The temperature estimate for the 3C58 central source was based
on an assumed distance of 8 kpc which was later reduced by a factor of 3
(Green \& Gull 1982): with this new distance, the resulting temperature
would be low enough to be marginally inconsistent with the ``standard''
cooling model.
	Moreover the age of 3C58 is based on an association with the
A.D.  1191 supernova, but Becker \etal (1982)
questioned this association, arguing from the low ratio of X-ray to radio
luminosities of the remnant that it is probably much older; this would
ruin its relevance for comparison with models of early cooling of
neutron stars.
	3) For the Crab pulsar (the only case of these three whose
existence and age are beyond doubt), the temperature estimate
is based on an upper
limit of the flux between the pulses of the X-ray curve observed by {\em
Einstein}: the pulsar is {\em undetected} at this phase, and thus the
temperature can hence be anywhere below the reported upper limit of $2.5
\times 10^6$ K.
	In none of these three cases was there any spectral evidence
that the X-ray emission is thermal emission from the surface of the star,
since the Crab and RCW 103 observations were done with the HRI detector,
which had no energy resolution, and the count rate from the 3C58 point
source in the IPC detector was too low to provide useful spectral
information.
	Moreover, the magnetospheric X-ray emission from such young
neutron stars is so strong that there is little hope of detecting
thermal radiation from the surface of the star itself (\"{O}gelman 1993).
	A fourth neutron star young enough to allow us to distinguish
between slow and fast neutrino cooling, and in this case with good data,
is PSR 0833-45 (Vela): comparisons with theoretical models have shown
repeatedly that its surface temperature is too low for ``standard''
cooling and requires a fast cooling agent (Nomoto \& Tsuruta 1986;
Shibazaki \& Lamb 1989; Page \& Baron 1990; Van Riper 1991; Page \&
Applegate 1992; Umeda {\em et al} 1992, 1993).
	It is therefore a reasonable theoretical {\em a priori} to interpret
all data within the fast neutrino cooling scenario.
	Nevertheless, as stated in section 3.1, we still consider
both types of neutrino cooling, letting observation be the ultimate
judge.

\subsection{PSR 0656+14 and PSR 1055-52}

	Since analyses of the {\em ROSAT} observations of PSR 0656+14 and PSR
1055-52 have been published recently, we will now briefly comment on these
results in light of the preceding remarks.
	The data are plotted in Figure~\ref{fig6}, along with some typical
cooling curves.

	In the case of PSR 0656+14, when the spectral fit is done with a
blackbody, the resulting surface temperature is $9.0 \pm 0.4 \times
10^5$ K, while a nonmagnetic helium atmosphere gives
$2.2 \pm 0.2 \times 10^5$ K (Finley \etal 1992).
	For a spin-down age of $1.1 \times 10^5$ yr the first value
is perfectly compatible with the slow neutrino cooling scenario
(without core pairing or with an appropriate proportion of core baryons
paired),
while the second is in disagreement and needs fast neutrino cooling,
unless the star is older than its spin-down age.
	An analysis with a magnetic hydrogen atmosphere spectrum gives
an intermediate value of $6.9^{+0.5}_{-0.3} \times 10^5$ K (Anderson
{\em et al.} 1993), which is lower than previously predicted for
slow neutrino cooling but slightly higher than our new results
in the case of complete pairing of the core:
	this temperature can easily be accomodated within this model by
having almost complete pairing of neutrons and protons in the core,
i.e. by slightly increasing the total specific heat compared to the completely
paired case.
	PSR 0656+14 would have to be much younger than its spin-down age
indicates for the magnetic temperature estimate to be incompatible with
the slow neutrino cooling scenario.

	For PSR 1055-52, spectral fits with a blackbody give a
surface temperature of $7.0 \pm 0.6 \times 10^5$ K (\"{O}gelman \&
Finley 1993).
	Given the spin-down age of $5 \times 10^5$ years,
\"{O}gelman \& Finley conclude that this temperature is compatible
with the slow neutrino cooling models:
	this is true only for models {\em without pairing in the
core} or with only protons paired.
	This temperature is slightly too high compared to slow
cooling whith neutron core pairing, but heating may explain
the discrepancy, and moreover, fits with non black-body spectra will
give lower temperatures and require less heating, if any at all.
	If both neutrons and protons are paired within the whole core,
the theoretical temperature with slow neutrino cooling is much lower
than the $7 \times 10^5$ K reported.
	If we compare this result with those from the heating models
of Shibazaki \& Lamb (1989) and Cheng {\em et al} (1992) only the
maximum heating rates of these models could justify the discrepancy in
this case:
	thus, {\em PSR 1055-52 most probably contains an unpaired
component in its core}.
	Being more speculative, if one adopts the idea that the
critical density for fast neutrino emission is low and that,
consequently, all neutron stars undergoing slow neutrino cooling have
their whole core paired and follow the trajectory of
Figure~\ref{fig2}d, then this reported temperature of $7 \times 10^5$
K is incompatible with the slow neutrino cooling scenario unless a
very efficient heating mechanism is at work.

\section{CONCLUSIONS}

	We have compared the recent temperature measurement of the
Geminga neutron star with cooling models and found that, since this star
is old enough to be in the photon cooling era, both fast and slow
neutrino emission mechanisms can explain its temperature.
	One therefore cannot draw any conclusion about neutrino emission
from dense nuclear matter using this observation alone.
	However, {\em a crucial feature in both types of models is that
they need nucleon pairing in most, if not all, of the core}.
	If no pairing is assumed in the core, then the predicted
temperature is either too high (slow neutrino cooling) or too low (fast
neutrino cooling) when compared to the observed temperature of Geminga.

	With fast neutrino cooling, nucleon pairing is needed to stop the
early cooling which, without this, would produce a star with a temperature
much lower than what is observed.
	If the fast neutrino emission is from hyperonic processes it is
possible that the suppression we observe is due to hyperon
superfluidity.
	It is not possible to distinguish between the various fast
processes, however; the theoretical uncertainty about $T_c$ allows us
to accomodate very different neutrino emission rates.
	Moreover, within a given fast cooling scenario, the observational
uncertainty also allows very different values of $T_c$.
	However, since fast neutrino emission occurs down to the very
center of the core, to be compatible with the Geminga observation,
these scenarios need pairing up to the highest density reached in this
object, with pairing critical temperatures higher than $10^9$ K.

	Within the slow neutrino cooling model (the ``standard'' model),
superfluidity is also needed, but for a different reason.
	The observed temperature is below what the simple model without
pairing predicts, but since this star is cooling by photon emission, we
can accelerate the cooling at this time by decreasing the specific heat
through pairing.
	The theoretical curves, in the photon cooling era, are very
insensitive to the high-density EOS or the star mass, the only
determining factor being how much of the core specific heat has been
eliminated by pairing.
	If we accept an age of $3 \times 10^5$ yr and a temperature
of $5 \times 10^5$ K, then the specific heat must have been reduced to
about 25\% of its normal value.
	This can be obtained either by pairing of the neutrons in the
whole core or by a combination of neutron and proton pairing, but
even in this case most of the neutrons must be paired.
	If both neutrons and protons are paired in the whole core, photon
cooling becomes so efficient that a substantial amount of heating is
needed, but several possible mechanisms have been proposed and may be
able to provide sufficient heating.

	The above discussion shows that, in order to distinguish clearly
between the fast and slow neutrino cooling scenarios, we need
observations of neutron stars younger than $5 \times 10^4$ yr, for which
the slow cooling scenario predicts temperatures higher than $0.9 - 1.1
\times 10^6$ K, depending on the exact age; at later times, both
scenarios can accomodate most observable temperatures depending on the
amount of pairing assumed.
	Geminga is old enough that the effect of the early neutrino
cooling has been washed out.
	However, our analysis showed that this star
does tell us -independently of its earlier neutrino cooling history-
that most, if not all, of its core is paired.
	PSR 0656+14 is at the limit at which we can still distinguish the
effect of fast neutino cooling, but the present uncertainty about its
temperature precludes drawing any conclusion.
	PSR 1055-52 can be also interpreted within both types of
neutrino scenarios, but its core must contain an unpaired component whose
specific heat keeps the star warm despite its age.
	If one accepts the fast neutrino cooling scenario as universal
(a reasonable theoretical {\em a priori}), then these three objects show
evidence of baryon pairing down to their very center, needed for
suppression of early neutrino emission, but none of them requires
this scenario.

\acknowledgments

I am grateful to T.  Ainsworth, J.  H.  Applegate, J.  P.  Halpern,
M.  Prakash, and M.  Ruderman for discussions.
This work was supported by HEA-NASA grant NAGW 3075 and, in its early
phases, by a fellowship from the Swiss National Science Foundation.
This work is contribution number 544 of the Columbia Astrophysics
Laboratory.

\clearpage 
\begin{table}[t]
\begin{center}
\begin{tabular}{ccc}
\tableline 
 Process Name & Process & $\begin{array}{c} {\rm Emissivity \;} Q_\nu \\
                                            {\rm (erg/sec/cm^3)}
                           \end{array}$  \\
\tableline 
a) Modified URCA
&
$ \left\{ \begin{array}{c}
                 n+n' \rightarrow n'+p+e^-+\overline{\nu_e} \\
                 n'+p+e^- \rightarrow n'+n+\nu_e
          \end{array} \right. $
&
$\sim 10^{20} \cdot T_9^8$  \\
b) K-condensate
&
$ \left\{ \begin{array}{c}
                 n+K^- \rightarrow n+e^-+\overline{\nu_e} \\
                 n+e^- \rightarrow n+K^-+\nu_e
          \end{array} \right. $
&
$\sim 10^{24} \cdot T_9^6$ \\
c) $\pi$ - condensate
&
$ \left\{ \begin{array}{c}
                 n+\pi^- \rightarrow n+e^-+\overline{\nu_e} \\
                 n+e^-   \rightarrow n+\pi^-+\nu_e
          \end{array} \right. $
&
$\sim 10^{26} \cdot T_9^6$ \\
d) Direct URCA
&
$ \left\{ \begin{array}{c}
                 n  \rightarrow p+e^-+\overline{\nu_e} \\
                 p+e^- \rightarrow n+\nu_e
          \end{array} \right. $
&
$\sim 10^{27} \cdot T_9^6$ \\
e) Quark URCA
&
$ \left\{ \begin{array}{c}
                 d \rightarrow u+e^-+\overline{\nu_e} \\
                 u+e^- \rightarrow d+\nu_e
          \end{array} \right. $
&
$\sim 10^{26} \alpha_c T_9^6$ \\
\end{tabular}
\end{center}
\caption[Neutrino Processes]
         {Some core neutrino emission processes and their emissivities.
          The emissivities are from : a) Friman \& Maxwell 1979,
          b) Brown {\em et al} 1988, c) Maxwell {\em et al} 1977,
          d) Lattimer {\em et al} 1991 and e) Iwamoto 1980.
          $T_9$ is the temperature in units of $10^9$ kelvins.
          \label{tab:nu}}
\end{table}

\clearpage 

\begin{table}[t]
\begin{center}
\begin{tabular}{ccccccc}
\tableline 
 EOS & \multicolumn{3}{c}{Maximum-Mass Star}
        & \multicolumn{3}{c}{$1.4 \; M_{\odot}$ Star}   \\

   .    & $M (M_{\odot})$ & $\rho_c (fm^{-3})$  & $x_p (\%)$
        &     $R (km)$          &    $\rho_c (fm^{-3})$   & $x_p  (\%)$     \\
\tableline 
  FP                    &      1.79       &     1.18     & $\sim 0$
                        &     10.85       &     0.69     &     2.0    \\

WFF(av14)               &     2.10       &     1.25     &      4.8
                        &     10.60       &     0.64     &     9.6    \\

  MPA                   &     2.44        &     0.89     &   19.0
                        &    12.45        &     0.41     &     9.0    \\

 PAL32                  &     1.68        &     1.51     &   15.2
                        &    11.02        &     0.74     &   11.4    \\

 PAL33                  &     1.90        &     1.24     &   14.0
                        &    11.91        &     0.54     &   9.7    \\
\tableline 
\end{tabular}
\end{center}
\caption[Properties of the EOS's used]
         {Some properties of the EOS's used for slow neutrino cooling.
          Columns two to four list the mass, central density and
          central proton fraction of a maximum mass star.
          Columns five to seven give properties of a
          $1.4 {\rm M_{\odot}}$ star :
          radius, central density and central proton fraction.
          (The PAL EOSs are labeled PALij, i,j=1,2,3, where i refers
          to the symmetry energy function and j to the compression modulus)
          \label{tab:EOS}}
\end{table}

\clearpage 

\begin{table}[t]
\begin{center}
\begin{tabular}{ccccrc}
\hline 
 EOS & \multicolumn{3}{c}{Core Components}
        & \multicolumn{2}{c}{Crust Components}   \\

   .    & n & p & e & n & e \\
\hline 
  FP      &   9.89 & 2.89 & 0.50 & 0.97 & 0.025 \\
WFF(av14) &   9.31 & 2.81 & 0.51 & 0.72 & 0.018 \\
  MPA     &  11.90 & 3.90 & 0.67 & 1.60 & 0.044 \\
 PAL32    &   9.66 & 3.41 & 0.68 & 1.16 & 0.031 \\
 PAL33    &  10.96 & 3.72 & 0.68 & 1.42 & 0.037 \\
\hline 
\end{tabular}
\end{center}
\caption[Specific heat]
         {Normal specific heat, at $T = 10^9$ K, of neutrons, protons and
          electrons in the core and crust of a $1.4$ \Msol neutron star built
          with the five EOS's used for slow neutrino cooling.
          (Units are $10^{38} \; ergs \; K^{-1}$, i.e.
          $C_v(T) = ({\rm Table-Entry}) \times (T/10^9 K) \times 10^{38} \;
          ergs \; K^{-1}$).
          \label{tab:Cv}}
\end{table}

\clearpage 


\clearpage 

\bigskip

\bigskip

\begin{figure}
\caption{\label{fig3}
         Cooling by the modified Urca process: effect of the star mass.
         EOS WFF(av14) and star mass of 1.2 \Msol (dash-dotted),
         1.4 \Msol (continuous), 1.6 \Msol (dash-triple dotted) and
         1.8 \Msol (dashed) with the same pairings as in Fig. 2.
         Crust neutrons are paired according to Ainsworth \etal (1989).
         The temperature in ordinate is the effective temperature
         ``at infinity,'' i.e., redshifted.
         The cross shows Geminga's temperature and age.}
\end{figure}

\bigskip

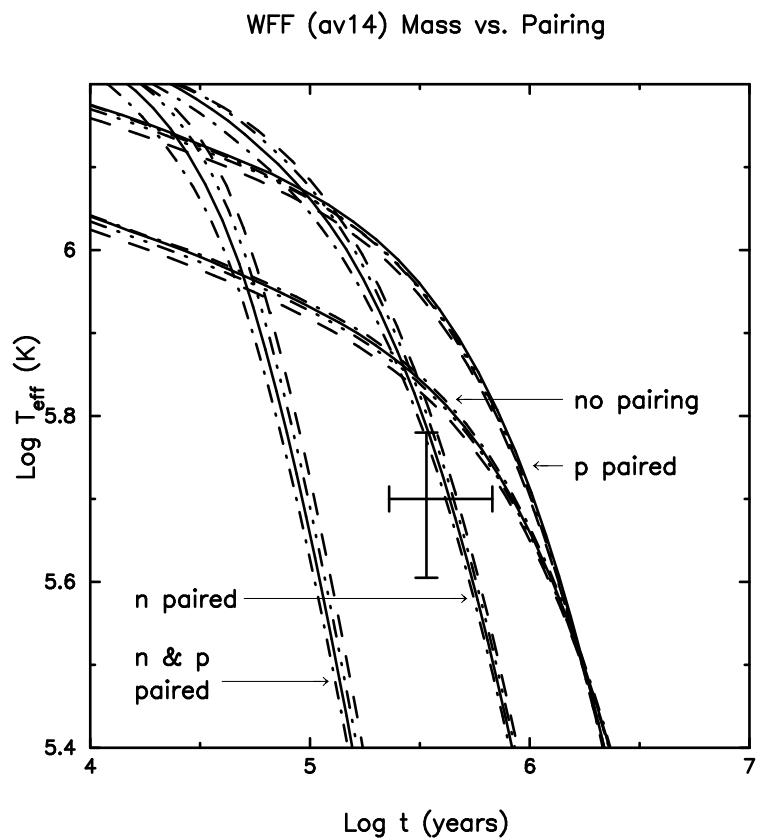
\begin{figure}
\caption{\label{fig4}
         Cooling by the modified Urca process: density dependent $T_c$.
         EOS WFF(av14) and star mass 1.4 \Msol.
         The continuous curve has no core neutron pairing,
         the other three curves have core neutron pairing
         as labeled (see Figure 1b).
         Core protons are not paired.
         Crust neutrons are paired according to Ainsworth \etal (1989).
         The temperature in ordinate is the effective temperature
         ``at infinity,'' i.e., redshifted.
         The cross shows Geminga's temperature and age.}
\end{figure}

\bigskip

\begin{figure}
\caption{\label{fig5}
         Fast neutrino cooling is not fast cooling forever:
	 the curves show the cooling history of two neutron stars of
         very different masses built with the same EOS, WFF(av14),
         and the same pairings (HGRR for core neutrons,
         CCY-PSi) for core protons, as labeled in Figure 1,
         and Ainsworth \etal 1989 for crust neutrons).
	 With this choice of pairings, neutrons are paired in the whole
         core for both stars, as are protons in the lighter star while
         the heavier star has a central region of unpaired protons.
	 We have assumed that a kaon condensate develops above a density
         of $10^{15} gm/cm^3$ such that the 1.6 \Msol star has a kaon ``pit''
         of 0.56 \Msol while the 1.0 \Msol star undergoes ``standard'' cooling.
	 During the neutrino cooling era
	 ($30 \; yr < age < 3 \times 10^4 yr$)
         the lighter star is warmer because of its low neutrino emission, while
         during the photon cooling era the heavier star is warmer because of
	 its larger specific heat provided by its unpaired protons.}
\end{figure}

\bigskip

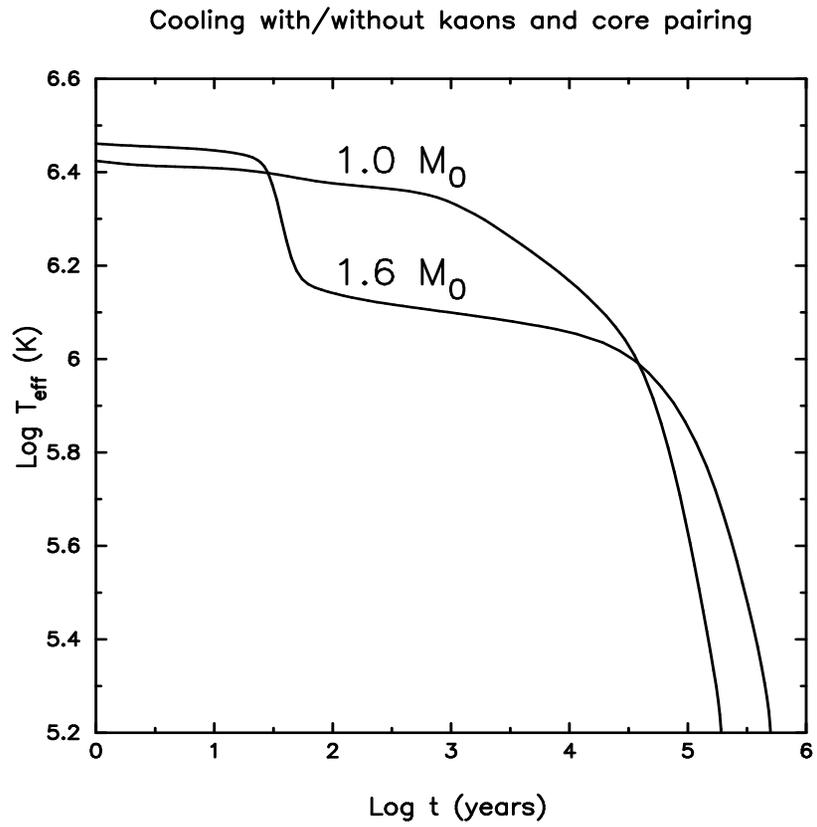
\begin{figure}
\caption{\label{fig6}
         Comparison of the estimated surface temperatures of PSR 0656+14
         and PSR 1055-52 with theoretical curves of slow neutrino cooling.
	 EOS WFF(av14) and star mass 1.4 \Msol, no pairing (continuous),
	 protons paired (dotted), neutrons paired (dash-dotted),
	 neutrons and protons paired (dash-triple-dotted).
         The three temperatures for PSR 0656+14 correspond to blackbody,
         magnetic hydrogen, and nonmagnetic helium atmospheres as indicated,
         while the PSR 1055-52 temperature is from a blackbody fit.
         The age ranges correspond to braking indices from 2 to 4, as we
         used for Geminga, the upper values probably being overestimates.}
\end{figure}

\clearpage 

\begin{flushleft}
\bf
Author Address
\end{flushleft}

\noindent
Dany Page: Instituto de Astronom\mbox{\'{\i}}a, U.N.A.M.,\\
           Apdo postal 70-264, 04510 MEXICO D.F.

\noindent
E-mail : PAGE@ASTROSCU.UNAM.MX\\

\end {document}
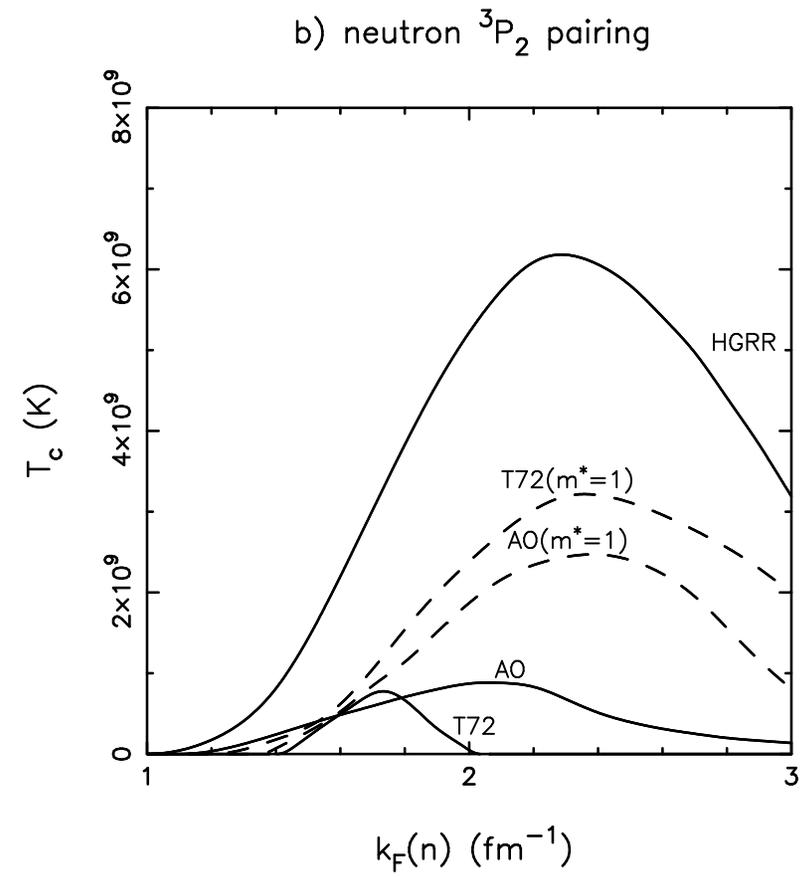
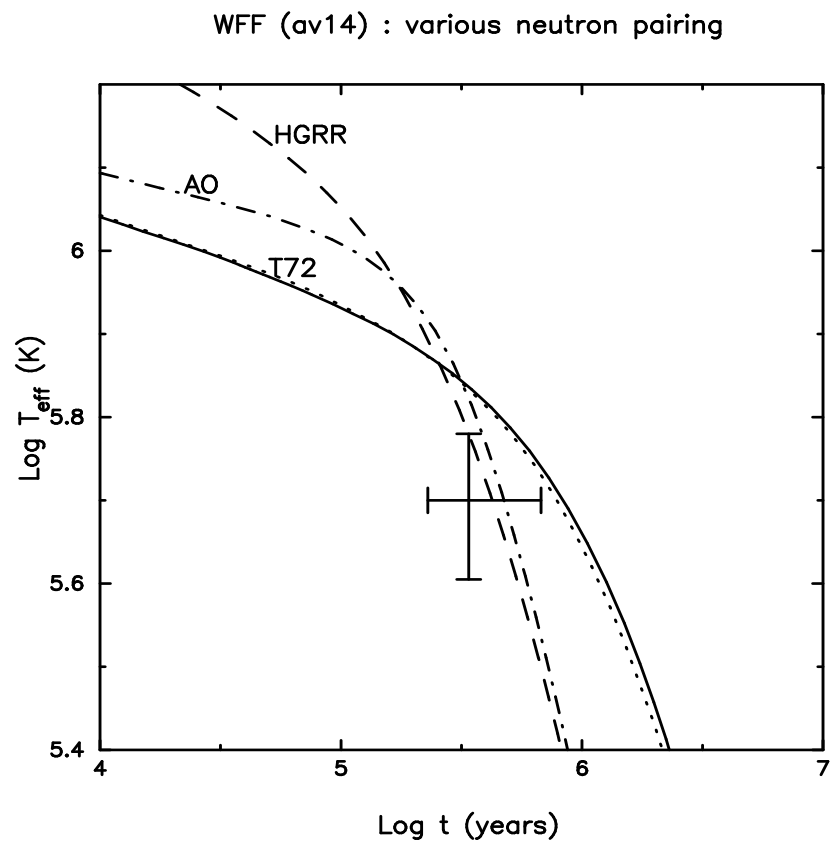
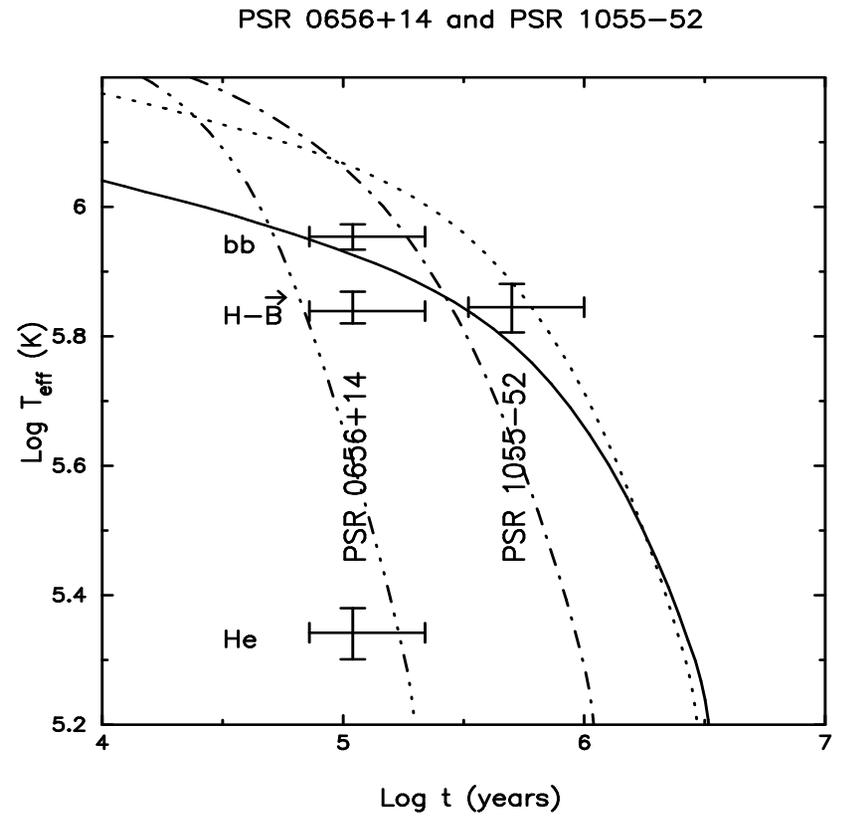
